# c-SELENE: Privacy-preserving Query Retrieval System on Heterogeneous Cloud Data

Diyah Puspitaningrum
Department of Computer Science,
University of Bengkulu
W.R. Supratman St., 38317 Bengkulu
Indonesia
diyahpuspitaningrum@gmail.com

## ABSTRACT

While working in collaborative team elsewhere sometimes the federated (huge) data are from heterogeneous cloud vendors. It is not only about the data privacy concern but also about how can those federated data can be querying from cloud directly in fast and securely way. Previous solution offered hybrid cloud between public and trusted private cloud. Another previous solution used encryption on MapReduce framework. But the challenge is we are working on heterogeneous clouds. In this paper, we present a novel technique for querying with privacy concern.

Since we take execution time into account, our basic idea is to use the data mining model by partitioning the federated databases in order to reduce the search and query time. By using model of the database it means we use only the summary or the very characteristic patterns of the database. Modeling is the Preserving Privacy Stage I, since by modeling the data is being symbolized. We implement encryption on the database as preserving privacy Stage II. Our system, called "c-SELENE" (stands for "cloud SELENE"), is designed to handle federated data on heterogeneous clouds: AWS, Microsoft Azure, and Google Cloud Platform with MapReduce technique.

In this paper we discuss preserving-privacy system and threat model, the format of federated data, the parallel programming (GPU programming and shared/memory systems), the parallel and secure algorithm for data mining model in distributed cloud, the cloud infrastructure/architecture, and the UIX design of the c-SELENE system. Other issues such as incremental method and the secure design of cloud architecture system (Virtual Machines across platform design) are still open to discuss. Our experiments should demonstrate the validity and practicality of the proposed high performance computing scheme.

## 1 INTRODUCTION

There are many cloud platforms available nowadays provide users capability to access, store, and sharing data. Having the data in a public cloud means they are vulnerable to the security attacks. Therefore we need a retrieval system that has capability of privacy-preserving that handle sensitive information while user query is retrieved on heterogeneous cloud data. Also a good retrieval system should be efficient in running time.

Previous naive solution for answering queries securely is by retrieving the encrypted database from the cloud to the client, then decrypt on the database, and then evaluate answer on the plain text database. But this approach consumes too much time thus impractical. Another solution consider to retrieve the $k$ nearest points for each point inside a dataset in a single query process. It is clear that the top-$k$ queries over encrypted databases reduce computation time rather than queries over plaintext databases. Vaidya [18] and Vaidya et al. [19] studied privacy-preserving top-$k$ queries in which the data are vertically partitioned. However, their solution only protects the query privacy. The data confidentiality and the access pattern are beyond the issues [3].

In the recent years multiple research papers focused around the distributed all $k$-nearest neighbor ((A)$k$NN) query processing on the cloud [12][13]. The naive approach of (A)$k$NN is to search for every point the whole dataset in order to estimate its $k$NN list. It is computationally very expensive. Since the cloud computing technology need infrastructure of managing distributed data among multiple servers, several researcher proposed the (A)$k$NN with MapReduce programming to process the large scale data efficiently.

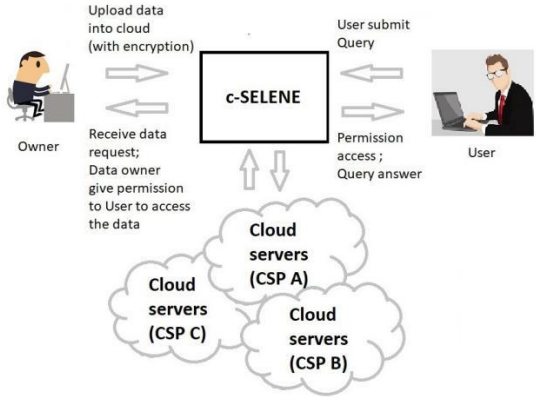

**Figure 1:** The c-SELENE in action

We interested to further research (A)$k$NN method on different distributed frameworks: i) one cloud, ii) distributed but similar platform clouds, and iii) heterogeneous clouds. All in GPU-based solutions. As benchmark, we compare our technique on two $k$NN-like queries viz. RNN (Reverse Nearest Neighbor) [8] and top-$k$ queries [18] [19].

Meanwhile, KRIMP [10], a state-of-the-art algorithm, finds interesting itemsets from a transaction database via the Minimum Description Length (MDL) principle. It finds the best set of patterns that describes the database best where the best set of itemsets is the set that provides the highest compression using MDL. Recent work [16] [4] [5] describes *Widening*, a framework for employing parallel resources to increase accuracy to find more interesting sets of itemsets than those found by the standard KRIMP algorithm. For Δ refining function, Widened KRIMP using *p-dispersion-min-sum* and *p-dispersion-sum* select maximally diverse subsets of candidates from the candidate table. The results show significantly better solutions than that of the traditional greedy algorithm, the standard KRIMP, by searching diverse regions of a solution space in parallel [16]. The pitfall of the Widened KRIMP is that both a performance evaluation ($\psi$) and a synchronized comparison of results from the parallel workers are required. KRIMP itself is actually an NB-like algorithm, thus we can parallel several refining set or models.

Here we proposed efficient method, c-ParMDL, for processing $k$ nearest neighbor for a specific user that made the query at the first place with privacy preserving parallel MDL model of database over encrypted data. To eliminate the pitfall of widening the standard KRIMP [16] we will investigate the use of careful pattern selection on itemsets through $k$NN while widening the standard KRIMP, and do pruning on the dataset compression thus improving to corresponding fruitful solution space. This algorithm is utilized on a user interface of a cross-cloud retrieval system called c-SELENE where user can query securely and quickly over databases from diverse cloud platforms.

The potential implementations of this research are: 1) to execute simple and advanced query, such as the one we propose in this paper; 2) another possibility of further development is to do data mining task such as searching similar information about some items lay on the relevant databases of documents across clouds (multi-CSP), by clustering on the set of databases (see our previous work of the MDL-based Frequent Itemset Hierarchical Clustering in [15][6]).

While using c-ParMDL to execute simple and complex query, start from the standard KRIMP, the pattern selection is carefully selected by the help of $k$NN method approach over cloud ($k$ points) as in [13]. One of the proposal of pruning frequent itemset candidates is by merging shorter code into longer codes with some predefined frequency. As a UIX interface system for querying similar data across CSP, Fig.1 shows a summary of how c-SELENE works.

In the rest of this paper, we first define the problem that we address. After that we briefly introduce our proposed approach, and then discuss the related work.

## 2 PROBLEM DEFINITION

In this section, we describe that the main goal of this research can be summarized in several items:

- The need of "federation" on experimental data. It is a multi-table mining problem.
  
  We need better methods for processing such multiple tables, without having to create a single large view. The federated data is collected through several worldwide archives distributed in heterogeneous platform clouds. It need standard format for access the data;

- The implementation of reliable computing instruments for data exploration, mining and knowledge extraction, user-friendly, scalable and as much as possible asynchronous (format/protocol) for querying.
  
  These can include: 1) the parallel data mining technique used for federated database modeling; 2) the UIX design of c-SELENE user interface integrated system; 3) the secure design of cloud architecture system (the design of virtual machines across diverse platforms) for experiment.

All the process in this research should demonstrates the effectiveness of the two basic requirements above.

## 3 APPROACH

The c-SELENE research aims at creating a prototype of querying integrated and asynchronous access system to federated data from diverse platform of clouds towards a user friendly, secure and time efficient retrieval system. We will test the system on both simple queries and complex queries. Overview of c-SELENE schema is on Fig. 2.

### 3.1 Preserving-Privacy System and Threat Model

Given four different entities: data owner, authorized user, private cloud, and public cloud. Owner has a collection of relation schema $S(X, A, F)$ to be stored in the heterogeneous clouds, where $X$ is object set $\{x_1, x_2, ..., x_n\}$; $A$ is attribute set $\{a_1, a_2, ..., a_m\}$; and $F$ is the relation set between $X$ and $A$, with $F = \{f_k: X \to V_k\}$. $V_k$ is element of $A$. Different with [20] that work on classification task, our proposed system used generic model of data mining, this includes the association rule mining model of a database. Our purpose is to answer either simple query or complex query on any database. Assume all clouds are located publicly, both for sensitive or insensitive data. c-SELENE has two privacy-preserving stages: first is by symbolized and extract only very characteristic of the database; second is after answering the query using model then the answer is encrypted using one of encryption algorithm and the answer is returned back to the user along with the encrypted code table symbols (see Fig. 2). Only the user that has key that can decrypt the query answer and understand the decrypted query answer through the code table reading.

## 3.2 Secure Cloud Database System

We do not view a cloud as hybrid cloud (consists of private and public cloud) such as in [20]. This simplicity will help to reduce time inefficiency. About the threat model, we assume Cloud Service Provider (CSP) is malicious or untrusted. The CSP could tamper data stored in cloud since the CSP has full control of the cloud; the CSP could search a subset of the data. To handle this we compute hashing of minimal data query unit (data fragment) before uploading data into clouds and also do meta data checking on every data fragments. The tampered data can be reconstructed by understanding the principle of MDL. For more privacy-preserving, $k$-anonymity is applied to sensitive data (Fig.3) [7].

| ZIPcode ($X$) | Age ($a_1$) | Disease ($a_2$) | ZIPcode ($X$) | Age ($a_1$) | Disease ($a_2$) |
|---|---|---|---|---|---|
| 17601 | 31 | Cancer | 176** | [30,35) | Cancer |
| 17601 | 32 | BRCA Mutation | 176** | [30,35) | BRCA Mutation |
| 17605 | 33 | Cancer | 176** | [30,35) | Cancer |
| 17605 | 34 | Alzheimer | 176** | [30,35) | BRCA Mutation |
| 13059 | 36 | Cancer | 1305* | [35,40) | Cancer |
| 13056 | 38 | Cancer | 1305* | [35,40) | Cancer |
| 13054 | 37 | Viral Infection | 1305* | [35,40) | Viral Infection |
| 13055 | 38 | Viral Infection | 1305* | [35,40) | Viral Infection |
| 13059 | 39 | Viral Infection | 1305* | [35,40) | Viral Infection |

**Figure 3:** A sample of $k$-anonymity ($k$=4, right). Left is the original table

From Fig.4 [2] all databases of database owner are encrypted. To query an outsourced database, the database owner must communicate through secure server as if the entire database were stored in the secure server. All encrypted database is partitioned, stored, and modeled distributedly within the secure cluster. The secue server do query processing using model algorithm (c-ParMDL) in the fragmented databases and return the encrypted model to the database owner. Query answer (QA) is generated once pruning-and-merging process is done results one single code table that then used to answer user query (Section 3.5.2).

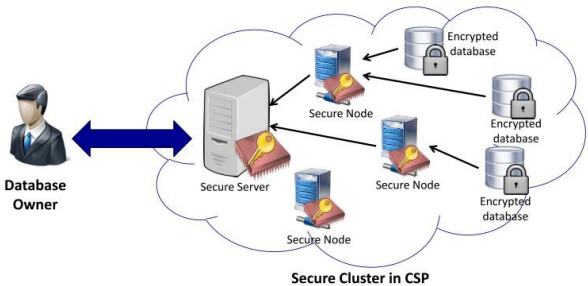

**Figure 4:** Secure cluster schema in each c-SELENE CSP

## 3.3 Format of Federated Data

Assume the databases used are located in heterogeneous clouds. A user query a complex query which involves database $DB = \{DB_1, DB_2, DB_3\}$, where each database lies on different clouds (different CSP). For fast processing each computation on fragments of all databases in $DB$ is processed parallel using MapReduce. Since we will use KRIMP model for association rules mining, we employ two type of layouts: horizontal or vertical database (Fig.5) [11] depend on the query operation (selection, projection, join, etc.).

| Transaction Id | Items |   | Item | Transaction Id Set |
|---|---|---|---|---|
| 1 | 2,1,5,3 |   | 1 | 1,3,4,5 |
| 2 | 2,3 |   | 2 | 1,2,5,6 |
| 3 | 1,4 |   | 3 | 1,2,4,5 |
| 4 | 3,1,5 |   | 4 | 3 |
| 5 | 2,1,3 |   | 5 | 1,4 |
| 6 | 2,4 |   |   |   |

**Figure 5:** Database layout: horizontal (left), vertical (right)

## 3.4 GPU Programming and Shared/Memory Systems

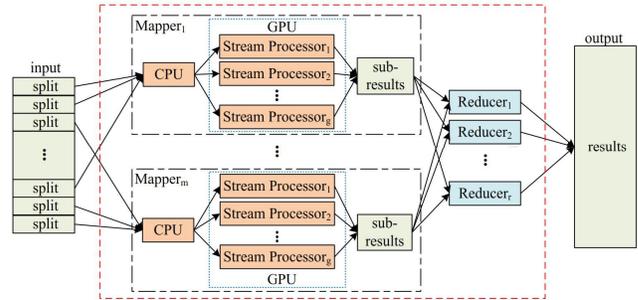

**Figure 6:** MapReduce model in heterogeneous architecture

We will use pyCUDA, Hadoop, and Spark in the cloud environment. Fig.6 shows parallel environment of the MDL based algorithm in multi-core techniques (GPU). Different with [1] that execute in one cloud, we employ three clouds, each from multiple CSP. On each cloud there are three machines with NVidia GPU installed. Per cloud, one machine serves as master, the two others are slaves. There is a local memory and shared memory through NVidia. For experiment we use very large synthetic datasets from Wikipedia (or others). To test the effect of memory capacity, the size of datasets is ranging from above 65, 128, 256, 512 (MB), and above 1, 2 (GB). Actually, we will test the datasets in four ways: 1) fragmented/distributed but in stand alone machine, 2) distributed but in one cloud and one CSP, 3) distributed, in multiple clouds, all in same CSP, 4) distributed, heterogeneous (in multiple clouds, and multiple CSP). See Fig.2. Our system should demonstrate the validity and practicality of the proposed high performance computing scheme.

## 3.5 Model Algorithm (c-ParMDL)

### 3.5.1 Standard Algorithm

Different from Apriori, in KRIMP the frequent itemsets are mined by picking frequent patterns that compress a database best (Fig.7). KRIMP finds the most interesting itemsets in a transaction database based on MDL, i.e., KRIMP defines the best model, $\mathcal{M}$, as the model that maximizes the compression of a transaction database, $\mathcal{D}$. Krimp calculates the size of the encoded database according to the Shannon entropy: $L(x) = -\log_2 P(x)$, where $L(x)$ is the code length measured in bits of an item or itemset in the database, and $P(x)$ is the relative frequency of the item's or itemset's appearance in the database [10][14]. The output KRIMP is *code table*, or *CT*, that is the summary of the interesting patterns in $\mathcal{D}$.

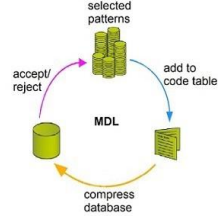

**Figure 7:** Activities in KRIMP standard

### 3.5.2 Parallel Algorithm

Widened KRIMP [16] introduces diversity of paths by *p-dispersion-min-sum* and *p-dispersion-sum*. They select maximally diverse subsets of candidates from the candidate table. *Directed placement* used for cover order of the *CT*. The heuristic function, *directed placement*, works by inserting the next candidate itemset, $F \in \mathcal{F}$, at a position with different fractional depths into $l$ parallel instances of *CT*. The depth inserted into *CT* is defined as $i/l\,|CT|$ where $i = 1, 2, …, l$. Since each itemset in the covering algorithm depends on its position in *CT*, positioning the next candidate itemset, $F$, at different depths results diverse paths.

As previously stated that the pitfall of the Widened KRIMP is that both a performance evaluation ($\psi$) and a synchronized comparison of results from the parallel workers are required. We propose different strategy. We proposed efficient $k$ nearest neighbor points approach called c-ParMDL, to process privacy preserving parallel MDL model on federated databases. After the distributed fragmented databases are securely processed in parallel, the frequent itemsets model then must be pruned and merged. Thus the model is short (compact) but lead to potential solution space.

**Algorithm 1** Privacy-Preserving c-ParMDL Algorithm
1: $L_1$ = {set of all singletons on $\mathcal{D}$}   // a singleton is a 1-itemset
2: **for** ($k=2; L_{k-1}\neq 0; k++$) **do**
3: $L_k = \varnothing$
4: $C_k$ = KRIMP($L_{k-1}$);   // KRIMP frequent itemset generation; the result is code table of the fragmented database
5: **for all** candidates $c \in C_k$ **do**
6:   **if** all the attributes in $c$ are entirely at any one party $P_l$ **then**
7:     party $P_l$ independently calculates $c.count$
8:   **else**
9:     let $P_l$ have $l_1$ of the attributes, …, $P_k$ have $l_k$ attributes ($\sum_{i=1}^{k} l_i = |c|$)
10:     construct $S_1$ on $P_1$'s side, …, $S_k$ on $P_k$'s side, // $S_k$ is a segmented dataset on party $P_k$
11:     where $S_i = S_{i1} \cap ... \cap S_{il_i}, 1 \le i \le k$
12:     compute $c.count = |\cap_{j=1..k} S_j|$ using Encryption Protocol
13:   **end if**
14:   $L_k = L_k \cup c|c.count \ge minsup$
15: **end for**
16: **end for**
17: $F = \cup_k L_k$
18: $F'$ = PruningMerging ($F$)   // replacing subset with the shortest parent code
19: QA = Do_Query_Operation($F'$,q)   // QA is the final answer.

### 3.5.3 Encryption Protocol

About the encryption protocol we use hashing where the sets of all the parties are hashed by all parties. The order of items is randomly permuted and the key known only by each party, the privacy is secured since no one except the party that can determine the mapping by previous party. For prevention, no intersection of two parties could be performed without knowing hashed values of set that belong to one party. Every party sends the remaining $c$ to its defined neighbor (i.e. left neighbor). The final result is taken from the cardinality of the total intersection set. If no collision means the result is correct. Else, if the collision value is below a predefined threshold it is still accepted. This situation may caused by same value of hash. See Fig.8 [2].

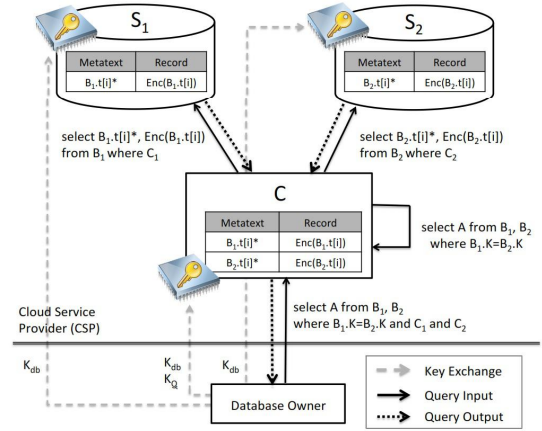

**Figure 8:** A schema of encryption on c-SELENE distributed database processing algorithm (c-ParMDL)

## 3.6 UIX Design

Fig.9 is a UIX design of the proposed system, the c-SELENE. It is simpler than [9] since we take execution time into account. In the *External Interfaces*, the SQL interface when executed implemented a user query and storage engines that translating the SQL requests to internal operations (to query wrapper and then after parsing turns into calling the databases). The location of the data is cached and new requests directly sent to the corresponding node. In *Partitioner*, The c-SELENE does databases partitioning for master/slaves or shared memory scenario. In *Protocol*, the secure protocol is tightly coupled with the parallel algorithm, c-ParMDL, to model the database. In *Encryption*, the resulted code table then encrypted for data secure purpose. In *Load Balancer*, the load balancer calculates load for each distributed node and allow user to adjust balance load between the distributed node processing by adding and removing nodes to the system. It is changing over time. The *MetaData* provides information about query data items (for example: keyID and data structure, databases name, tables, and storages). *Decryption* is used to check, accept, and reject the encrypted database/fragmented database requests. The final query answer is the output of a query operation on the distributed databases that create one single code table using c-ParMDL algorithm.

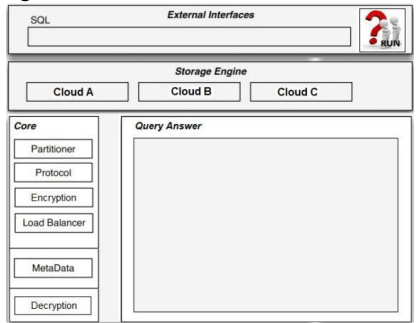

**Figure 9:** UIX design of c-SELENE

## 3.7 Experimental Design

The query operations performed on the privacy-preserving c-ParMDL algorithm (Section 3.5.2, Step 19) are: 1) simple query (projection, selection), or 2) advanced query (join/multiple join) as further development of our work in [17]. About multi-table mining, we do allow creating a single large view but we generate it from the distributed database models. Since a model is an approximation of the database, thus in our hypothesis the query execution on distributed model will be faster.

## 4 DISCUSSION

Query wrapper and RDBMS wrapper are topics that can be explored or discussed further [21]. Also here is a list of potential future research derived from this research:

1) *incremental methods*. The system must handle updates of new data and deletions without having to recompute patterns over the entire database; Given an updating on the database, $\mathcal{D}'$, with an additional new object set or item, $X'$, there is an algorithm for incremental weighted itemset cover that achieves the optimal aggregate value on every instance. Also For every incremental cover there is a reduced cover of no higher cost for any cover from previously available item;

2) *the secure design of cloud architecture system* (*Virtual Machines across platform design*). Beside the technical aspect of security in cloud system, for data reconstruction purpose tampered data after attack actually can be reconstructed/reapproximate using MDL principle.

## 5 CONCLUSIONS

In this paper, we have proposed a system to query integrated and asynchronous access system to federated data from heterogeneous clouds from different cloud server platforms towards a user friendly, secure and time efficient retrieval system. We have discussed issues about: preserving-privacy system and threat model, the format of federated data, the parallel programming (GPU programming and shared/memory systems), the parallel and secure algorithm for data mining model in distributed cloud, the cloud infrastructure/architecture, and the UIX design of the c-SELENE system.

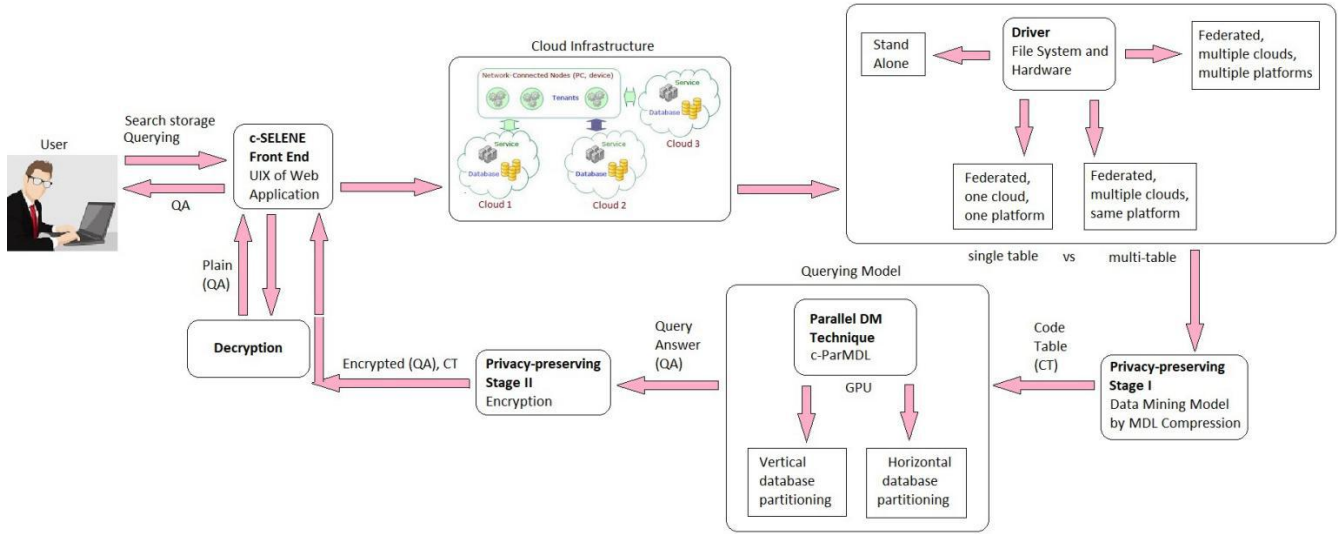

**Figure 2:** The architectural concept of c-SELENE, an integrated query engine across heterogeneous cloud platforms